\documentclass[twocolumn,aps,showpacs,prb,tightenlines,amsmath,amssymb,superscriptaddress]{revtex4}
\usepackage{bm}
\usepackage{ulem}
\usepackage{graphicx}
\usepackage{bmpsize}
\usepackage{amssymb}
\usepackage{amsmath}
\usepackage{colordvi}
\usepackage{calrsfs}

\begin{document}

\title{Coherent manipulation of a Majorana qubit by a mechanical resonator}

\author{P. Zhang}
\affiliation{CEMS, RIKEN, Saitama 351-0198, Japan}
\author{Franco Nori}
\affiliation{CEMS, RIKEN, Saitama 351-0198, Japan}
\affiliation{Department of Physics, The University of Michigan, Ann Arbor, Michigan 48109-1040, USA}
\date{\today}

\begin{abstract}
 
  We propose a hybrid system composed of a Majorana qubit and a nanomechanical
  resonator, implemented by a spin-orbit-coupled superconducting nanowire, using
  a set of static and oscillating ferromagnetic gates. The ferromagnetic gates
  induce Majorana bound states in the nanowire, which hybridize and constitute a Majorana qubit. Due to the
  oscillation of one of these gates, the Majorana qubit can be coherently rotated. By tuning the gate voltage to
  modulate the local spin-orbit coupling, it is possible to reach the resonance of the qubit-oscillator system for relatively strong couplings.
    
\end{abstract}
\maketitle

\section{Introduction}
Majorana bound states (MBS)\cite{Kitaev131,Beenakker113} are recently attracting increasing interest both
theoretically and experimentally. These have been predicted to exist in
artificial structures, such as nanowires with spin-orbit coupling (SOC) in proximity to a superconductor,\cite{Lutchyn077001,Oreg177002} ferromagnetic
atom chains on top of a superconductor,\cite{Choy195442} topological
insulator/superconductor hybrid
structures,\cite{fu096407,Rakhmanov075141,cook201105,Akzyanov085409,Akzyanov-arxiv,jiang075438} and
superconducting circuits.\cite{you5535} Recently, possible signatures of MBS have
been reported in nanowires,\cite{Mourik1003,das887,deng6414} atom chains,\cite{Nadj-Perg602} and topological insulator/superconductor structures.\cite{Xu017001}
The MBS attract considerable attention partly due to their hypothetical non-abelian
anyonic statistics, which might allow the realization of topologically-protected quantum
information manipulation.\cite{Beenakker113,nayak1083,Alicea412,Tewari010506} In
parallel to the ongoing search of some unambiguous confirmation of
MBS,\cite{Dumitrescu094505,Liu267002} there are also numerous theoretical
studies on how to efficiently exploit these MBS.

One promising application of MBS is to construct Majorana
qubits.\cite{nayak1083}  It has been suggested that Majorana qubits might be robust against local perturbations and are
  hence promising to store quantum information.\cite{nayak1083,mao174506,sau964} (Note that Majorana qubits are not totally protected from decoherence, as
studied in, e.g., Refs.~\onlinecite{rainis174533,schmidt085414,Goldstein205109,Budich121405}.) Furthermore, Majorana qubits could be
rotated by topologically-protected braiding operations.\cite{Alicea412,Ivanov268} Therefore,
among various realizations of
qubits,\cite{wuReview,perge2010,zhang115417,li086805,buluta326,buluta74,you474}
Majorana qubits are considered to be promising candidates for building blocks of quantum information
processors. The braiding operations alone are insufficient to
realize a universal quantum gate based on a Majorana qubit.\cite{nayak1083} For the implementation of arbitrary qubit rotations, other non-topological
operations are required. Several schemes of such non-topological operations
assisted by, e.g., phase gates,\cite{Bonderson180505,Clarke180519} quantum
dots,\cite{Flensberg090503,Bonderson130505} flux
qubits,\cite{Pekker107007,Hassler125002} or microwave cavities,\cite{Schmidt107006} have been proposed in the
literature.

Nanomechanical resonators\cite{Craighead1532} could also be used to study non-topological operations of a Majorana qubit. For example, quite
 recently, Kovalev {\sl et al.}\cite{kovaklev106402}  have proposed to rotate a Majorana qubit by a magnetic cantilever.
Indeed, nanomechanical resonators have been utilized to couple to a wide range of quantum systems, including electric
circuits,\cite{Blencowe159} optomechanical devices,\cite{Aspelmeyer1391}
atoms,\cite{Hammerer063005} Cooper-pair boxes,\cite{irish155311} spin 
qubits,\cite{rabl041302} or microwave cavities.\cite{arxiv1503.02437} With
the assistance of nanomechanical resonators, it is possible to perform important
applications such as quantum manipulations, quantum measurements, as well as
efficient sensing. These applications exploit the advantages of
nanomechanical resonators, e.g., their large quality factors ($10^3$-$10^6$), high natural frequencies
(MHz-GHz), as well as the feasibility of reaching the quantum ground states by cooling
methods.\cite{connell697,teufel359,chan89} Recently, nanomechanical resonators have also been exploited to measure or manipulate the
MBS.\cite{Walter224510,walter155431,chen050301}  Nevertheless, the study of hybrid systems\cite{xiang623} coupling
  nanomechanical resonators to Majorana qubits is  quite limited. 
This work aims to contribute to this field. In this paper, we propose another Majorana
qubit-nanomechanical resonator hybrid system in the framework of the spin-boson model,\cite{Leggett725} based on a semiconductor nanowire in
proximity to an s-wave superconductor. We show that a
strong coupling between a nanomechanical resonator and a Majorana
qubit can be achieved, allowing an efficient transfer of quantum information between these two
quantum systems. Further, with braiding operations, it should be possible to realize a universal quantum gate based on
a Majorana qubit.

This paper is organized as follows. First, we describe the Majorana qubit and its coupling to a nanomechanical resonator. Afterwards, we
numerically study the coupling strength and the resonance condition of the
hybrid system. Then, we solve the qubit-phonon dynamics and
achieve a coherent control of the Majorana qubit. Finally, we summarize our results.

\section{Model and Hamiltonian}
As illustrated in Fig.~\ref{fig1}, we consider a semiconductor nanowire with a
Rashba SOC of strength $\alpha^{\rm R}_0$ on the surface of an s-wave superconductor with a superconducting gap
$\Delta$. Three ferromagnetic gates, FM1, FM2 and FM3, are placed on top of and
along the nanowire. Among these gates, FM1 and FM3 are static while FM2 is free to
harmonically oscillate along the nanowire (with a mass $M$ and an oscillation frequency
$\omega_0$). The gates FM1 and FM3 are sufficiently long (of the order of
1-10~$\mu$m) while FM2 in between is relatively short (of the order of 100~nm). These ferromagnetic gates induce a local Zeeman splitting in the nanowire. For simplicity, we take the
Zeeman splitting under the three gates to be
identical, with a magnitude of $B_0$. An electric voltage $V$ can be applied on
the gates to modulate the Rashba SOC locally, e.g., from $\alpha^{\rm R}_0$ to $\alpha^{\rm
  R}_V$. In our study, we consider the case with $B_0^2>(\Delta^2+\mu^2)$, where $\mu$ is the chemical potential in the
nanowire. Therefore in the nanowire, the parts subject to the Zeeman splitting
(under the three ferromagnetic gates) are in the topological ($T$) region. The
remaining parts, without the Zeeman splitting,  are in the non-topological ($N$)
region.\cite{Oreg177002,Alicea412} As a result, the nanowire has an
$N$-$T$-$N$-$T$-$N$-$T$-$N$ domain structure where the three $T$ domains are
under the gates. At the six boundaries between the $N$ and $T$ domains in the nanowire, MBS arise. As the two outer MBS are far
apart, only the four inner ones, schematically labeled as $\gamma_1$-$\gamma_4$
in Fig.~\ref{fig1}, are coupled due to hybridization arising from their small
separation\cite{Bolech237002,Semenoff1479,Tewari027001,Kraus267002} and are hence relevant to our consideration.

\begin{figure}[hbt]
  {\includegraphics[width=8.5cm,height=4.6 cm]{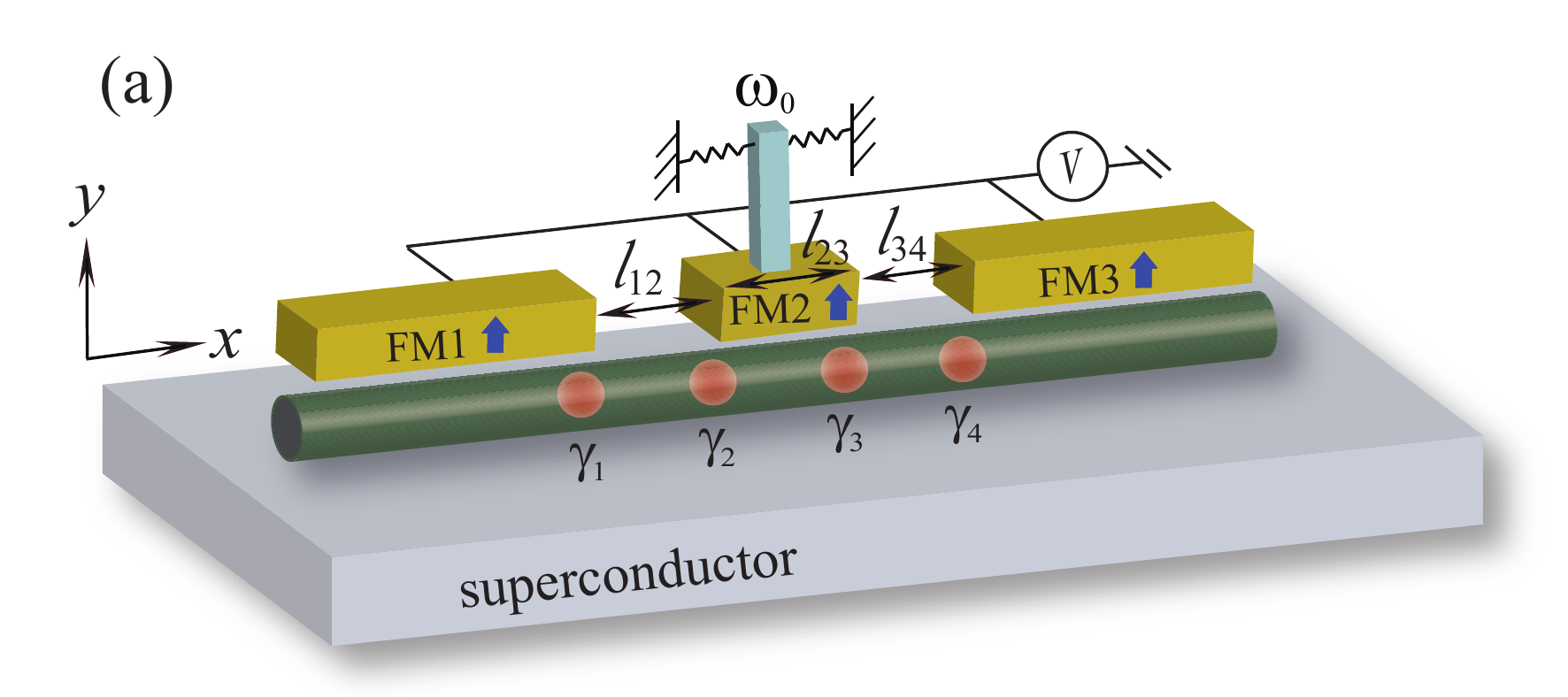}}\\
  {\includegraphics[width=8.5cm,height=5.2 cm]{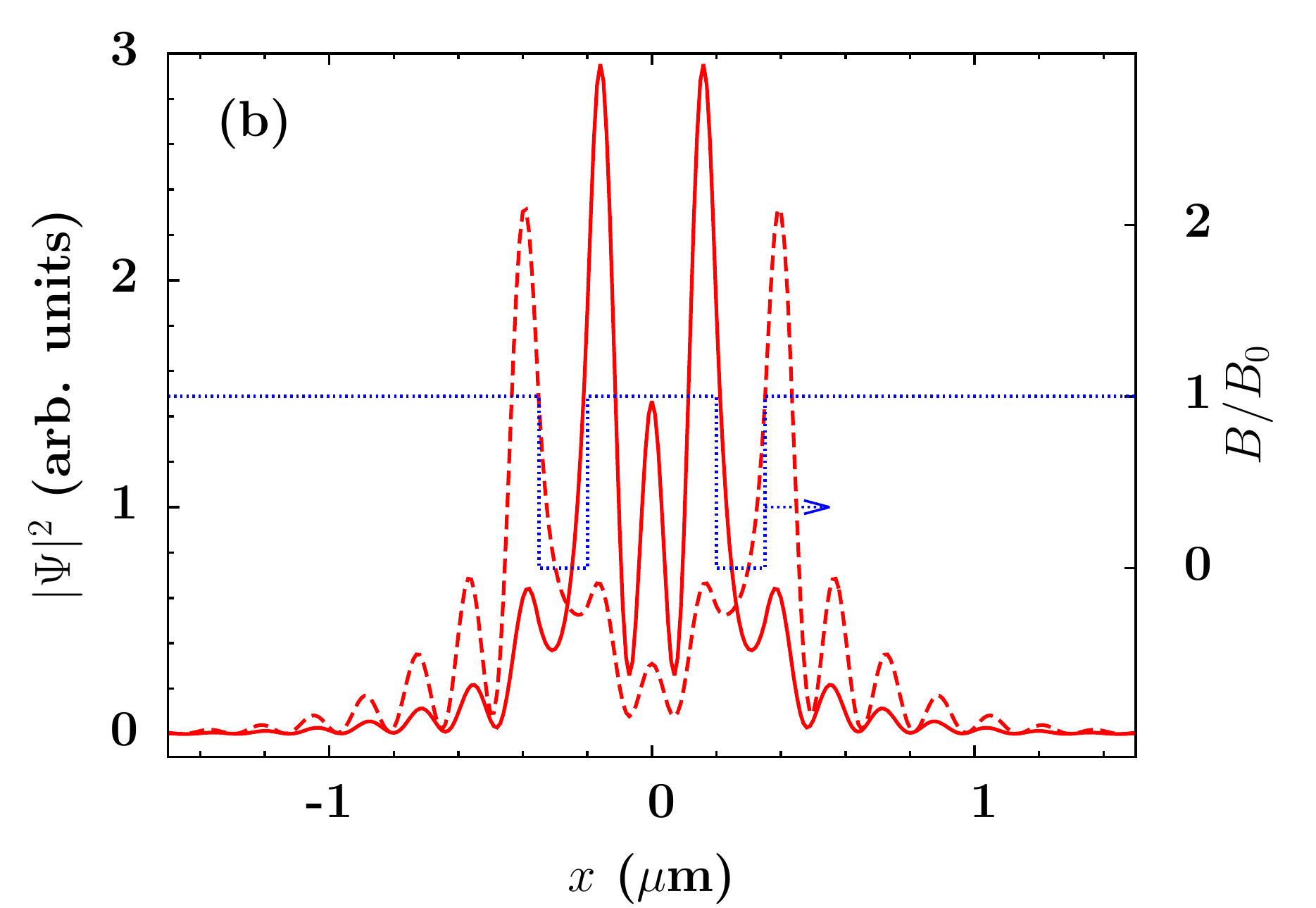}}
  \label{fig1}
  \caption{(Color online) (a) Schematic diagram of the proposed Majorana qubit-nanomechanical
    resonator hybrid system. A semiconductor nanowire is placed on the surface of an
    s-wave superconductor. Three ferromagnetic gates are on  top of the
    nanowire, of which FM1 and FM3 are sufficiently long (of the order of 1-10~$\mu$m) and static, while FM2 is
    relatively short (of the order of 100~nm) and free to oscillate as a harmonic oscillator. The
    ferromagnetic gates induce a local Zeeman splitting $B_0$ in the underlying
    nanowire, and can also be used to modulate the local Rashba SOC
    strength by applying an electric voltage $V$. (b) Wave amplitude $|\Psi|^2$ of the four
    coupled MBS at a static state in an InSb nanowire with the set-up shown in
    (a). The red dashed and red solid curves respectively correspond to the wave amplitudes of the
    lowest two eigenstates (close to the zero energy). These two states
    constitute the Majorana qubit. The dotted curve with the
    scale on the right-hand side of the frame indicates the profile of the inhomogeneous Zeeman
    splitting along the nanowire. In the calculation $l_{12}=l_{34}=150$~nm,
    $l_{23}=400$~nm, $B_0=1$~meV, and $\Delta=0.5$~meV. The gate voltage $V$ is
    zero and the Rashba SOC is homogeneous along the nanowire, with a strength
    $\alpha^{\rm R}_0=20$~meV nm.}
  \label{fig1}
\end{figure}

To lowest order, the hybrid system constructed above can be described by the Hamiltonian
\begin{align}
  H= H_{\rm M}+H_{\rm osc},
  \label{hm}
\end{align}
where the mutual coupling Hamiltonian of the MBS\cite{Bolech237002,Semenoff1479,Tewari027001,Kraus267002}

\begin{align}
  {H}_{\rm M}=ig_{n}[l_{12}(t)]\gamma_1\gamma_2+ig_{t}(l_{23})\gamma_2\gamma_3+ig_{n}[l_{34}(t)]\gamma_3\gamma_4,\label{hm}
\end{align}
and the nanomechanical oscillator Hamiltonian
\begin{align}
  { H}_{\rm osc}=\frac{p^2}{2M}+\frac{1}{2}M\omega_0^2x^2_0(t).
\end{align}
The coupling strengths $g_{n,t}$ depend on the domain lengths $l_{ij}$. Due to the
oscillation of the gate FM2, $l_{12}(t)=l_{12}^0+x_0(t)$ and  $l_{34}(t)=l_{34}^0-
x_0(t)$ are time dependent. Here, $x_0(t)$ stands for the displacement of the gate FM2 from its
balance position, which is much smaller than the static domain lengths
$l^0_{12,34}$. Therefore, to first order in $x_0$, one has
\begin{align}\nonumber
{ H}_{\rm M}=&i[g_n(l_{12}^0)+x_0(t)  g_{n}^\prime(l_{12}^0)]\gamma_1\gamma_2+ig_t(l_{23})\gamma_2\gamma_3\\&+i[g_n(l_{34}^0)-x_0(t)g^\prime_n(l_{34}^0)]\gamma_3\gamma_4.
\end{align}

The four MBS, satisfying $\{\gamma_i,\gamma_j\}=\delta_{ij}$, can be used to
construct a Majorana qubit as follows.\cite{nayak1083,Beenakker113} At first we define two Dirac fermion operators\cite{note}
$c_\uparrow=(\gamma_1+i\gamma_4)/\sqrt{2}$ and
$c_\downarrow=(\gamma_2+i\gamma_3)/\sqrt{2}$. The Hilbert space of ${ H}_{\rm M}$ can
then be spanned by states $|n_\uparrow,n_\downarrow\rangle$, with the fermion occupation numbers $n_\uparrow=c_\uparrow^\dagger
c_\uparrow$ and $n_\downarrow=c_\downarrow^\dagger
c_\downarrow$. Due to the conservation of fermion parity,
the states $\{|0,1\rangle,|1,0\rangle\}$ and $\{|0,0\rangle,|1,1\rangle\}$ form
two decoupled (odd and even)
sectors.\cite{Beenakker113,Schmidt107006,Pekker107007} We assume that there
is no high-energy excitation (e.g., no Cooper-pair breaking in the
superconducting substrate) and restrict our study to the odd sector
with $n_\uparrow+n_\downarrow=1$. For convenience, we define pseudo-spins
  $|\uparrow\rangle=|1,0\rangle$ and $|\downarrow\rangle=|0,1\rangle$, and use
them as the two logical states of the Majorana
qubit.\cite{Schmidt107006,Pekker107007,Flensberg090503,kovaklev106402} In this
pseudo-spin space, $i\gamma_1\gamma_2=-i\gamma_3\gamma_4=-\sigma_y$ and
$i\gamma_2\gamma_3=-\sigma_z$. The nanomechanical oscillator is quantized in the
Fock space $\{|n\rangle\}$ with $|n\rangle=\frac{(a^\dagger)^n}{\sqrt{n!}}
|0\rangle$, where $a=\sqrt{\frac{M\omega_0}{2\hbar}}(x_0+\frac{i}{M\omega_0}p)$
is the annihilation operator of phonons. Consequently, in the space $\{|\uparrow \downarrow\rangle\otimes |n\rangle\}$,
the hybrid system can be simply described by the spin-boson Hamiltonian,\cite{Leggett725}
\begin{align}
{ H}_{\rm eff}=-\frac{\varepsilon}{2}\sigma_z-\delta\sigma_y+g(a^\dagger
+a)\sigma_y+n\hbar\omega_0,
\label{he}
\end{align}
with the constant omitted. Here $\varepsilon=2g_t(l_{23})$, $\delta=g_n(l_{12}^0)-g_n(l_{34}^0)$, and
$g=-{\tilde x_0}[g_n^\prime (l_{12}^0)+g_n^\prime (l_{34}^0)]$, where ${\tilde
  x_0}=[\hbar/(2M\omega_0)]^{1/2}$ is the zero-point motion of the
oscillator.

\section{Hybridization of Majorana Bound States}
In this section, we study the MBS and their mutual coupling. In the static state, the
inhomogeneous nanowire can be described by a tight-binding model. Using the
Bogoliubov-de Gennes basis
$\Psi_j=(f_{j\uparrow},f_{j\downarrow},f_{j\downarrow}^\dagger,-f_{j\uparrow}^\dagger)$, where $f_{j\eta}$ stands for the fermion operator of a spin-$\eta$
($\eta=\uparrow$, $\downarrow$) electron on the $j$-th lattice site, the particle-hole Hamiltonian reads\cite{Choy195442}
\begin{align}
  H_{\rm BDG}=\frac{1}{2}\sum_j[\Psi_j^\dagger {\hat h}_j\Psi_j+(\Psi_j^\dagger {\hat t}_j\Psi_{j+1}+{\rm
      H.c.})],
  \label{hami}
\end{align}
where
\begin{align}
&{\hat h}_j=(2t_0-\mu)s_0\tau_z+\Delta s_0\tau_x+B_js_y\tau_0,\\
&{\hat t}_j=t_0s_0\tau_z+i\alpha_j s_z\tau_z.
\end{align}
In the above Hamiltonian, the Pauli matrices $\tau_{x,y,z}$ act on the particle-hole
space and $s_{x,y,z}$ act on the real spin space. The spin-diagonal hopping
energy is $t_0=\hbar^2/(2m^\ast a^2)$, and the spin-off-diagonal hopping
energy is $\alpha_j= \alpha^{\rm R}_j/(2a)$. Here $B_j$ and $\alpha^{\rm R}_j$ are
the on-site Zeeman splitting and Rashba SOC, respectively, $m^\ast$ is the effective
electron mass, and $a$ is the lattice spacing in the discretized
tight-binding model. In the $T$ ($N$) domains $B_j=B_0$ ($B_j=0$) and
$\alpha^{\rm R}_j=\alpha^{\rm R}_V$ ($\alpha^{\rm R}_j=\alpha^{\rm R}_0$). When
the gate voltage $V$ is zero, $\alpha^{\rm R}_V=\alpha^{\rm R}_0$.

Here, to lowest order, we follow Refs.~\onlinecite{kovaklev106402} and
\onlinecite{jiang075438} to investigate the coupling strength $g_n$ ($g_t$)
approximately in an isolated $T$-$N$-$T$
($N$-$T$-$N$) three-domain structure. In such a simplified model, the inner $N$
($T$) domain has a finite length, while the outer two $T$ ($N$) domains are assumed
to be infinitely long. By numerically diagonalizing this three-domain system, one can
obtain the energy splitting of the two MBS localized at the two $T$/$N$ boundaries. This energy splitting is precisely caused by
the coupling of the MBS. With $g_n$ and $g_t$ known numerically, the Majorana qubit can be well
described by $\varepsilon$ and $\delta$, and the qubit-phonon
coupling $g$ can be obtained also from $g_n^\prime$ [refer to Eq.~(\ref{he})]. Moreover, by exactly diagonalizing the Hamiltonian of the genuine
$N$-$T$-$N$-$T$-$N$-$T$-$N$ domain structure as shown in Fig.~\ref{fig1}(a), one can obtain the hybrid four
MBS under consideration.

\begin{figure}[hbt]
  {\includegraphics[width=8.5cm]{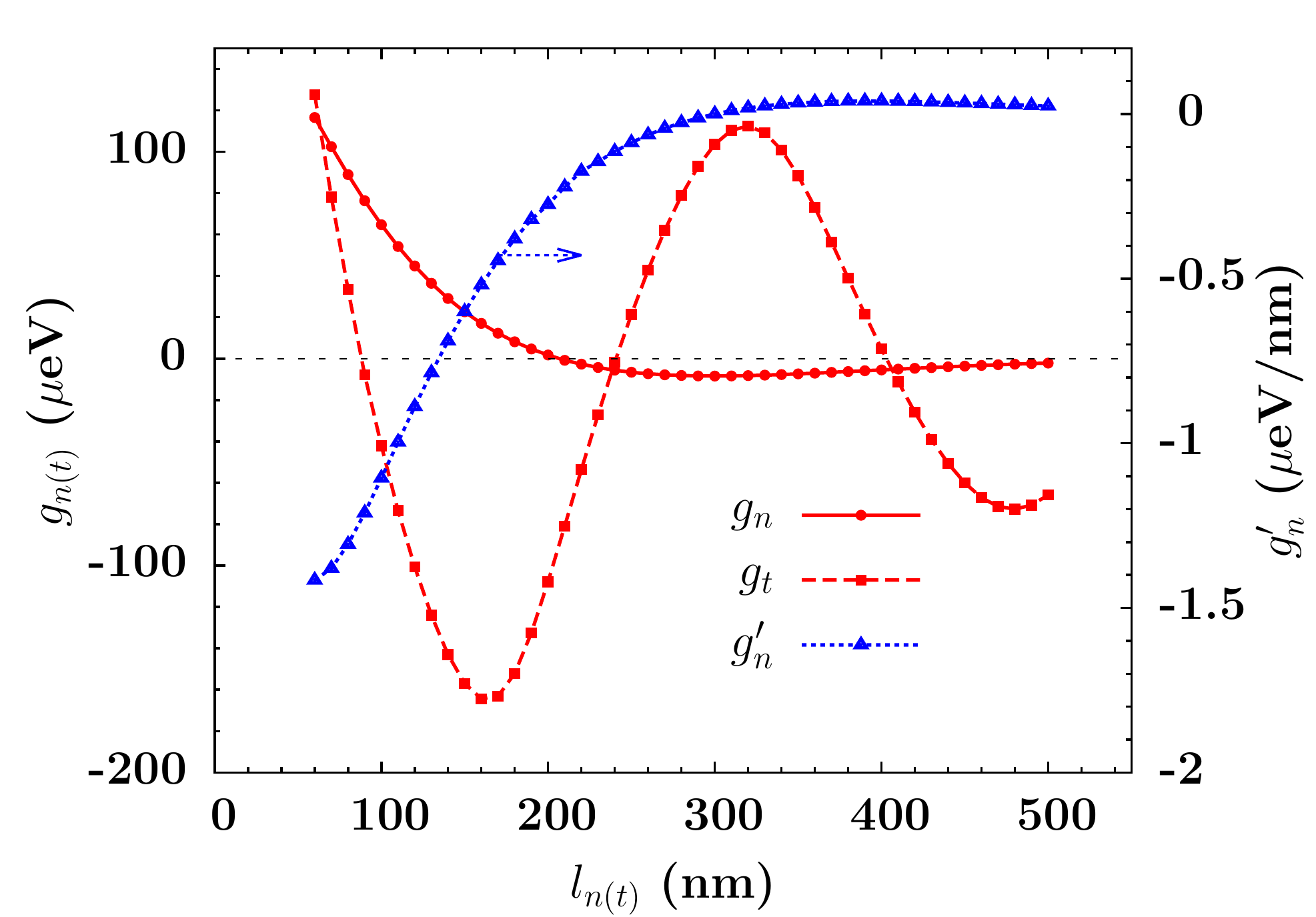}}
  \caption{(Color online) Majorana coupling strength $g_n$ ($g_t$) versus
    $l_n$ ($l_t$), the length of the inner $N$ ($T$) domain between the two outer
    $T$ ($N$) domains. The derivative
    $g_n^\prime$ versus $l_n$ is also shown, with the scale on
    the right hand side of the frame. The necessary parameters for the calculation
    are specified in the main text.}
  \label{fig2}
\end{figure}

In this work, we consider an InSb quantum wire\cite{Mourik1003} with an effective electron
mass $m^\ast=0.015~m_e$, a Rashba SOC $\alpha^{\rm  R}_0=20$~meV~nm, and a large
Landau factor $g_L\approx 50$. We choose the superconducting gap $\Delta=0.5$~meV, the
local Zeeman splitting $B_0=1$~meV, the chemical potential $\mu=0$, and the
lattice constant $a=10$~nm. The total lattice site number is chosen as 1000 for
the numerical convergence. In Fig.~\ref{fig2}, we show the dependence of
the Majorana coupling strength $g_n$ ($g_t$) on the length of the  $N$ ($T$)
domain $l_n$ ($l_t$), as well as the derivative $g_n^\prime$ versus
$l_n$. Further, as an example, in Fig.~\ref{fig1}(b) we present the wave amplitude $|\Psi|^2$ of the four
hybrid MBS, when $l_{12}=l_{34}=150$~nm and $l_{23}=400$~nm. In
Fig.~\ref{fig1}(b), the red dashed and red solid curves stand for the wave amplitudes of the
lowest two eigenstates (close to the zero energy) in the static inhomogeneous nanowire. The state corresponding to the red solid (dashed) curve is mainly
contributed by the $\gamma_2$ and $\gamma_3$ ($\gamma_1$ and $\gamma_4$) MBS. Here, these two states form the Majorana qubit.

\section{Qubit-phonon coupling and resonance}
\begin{figure}[thb]
  {\includegraphics[width=8.5cm]{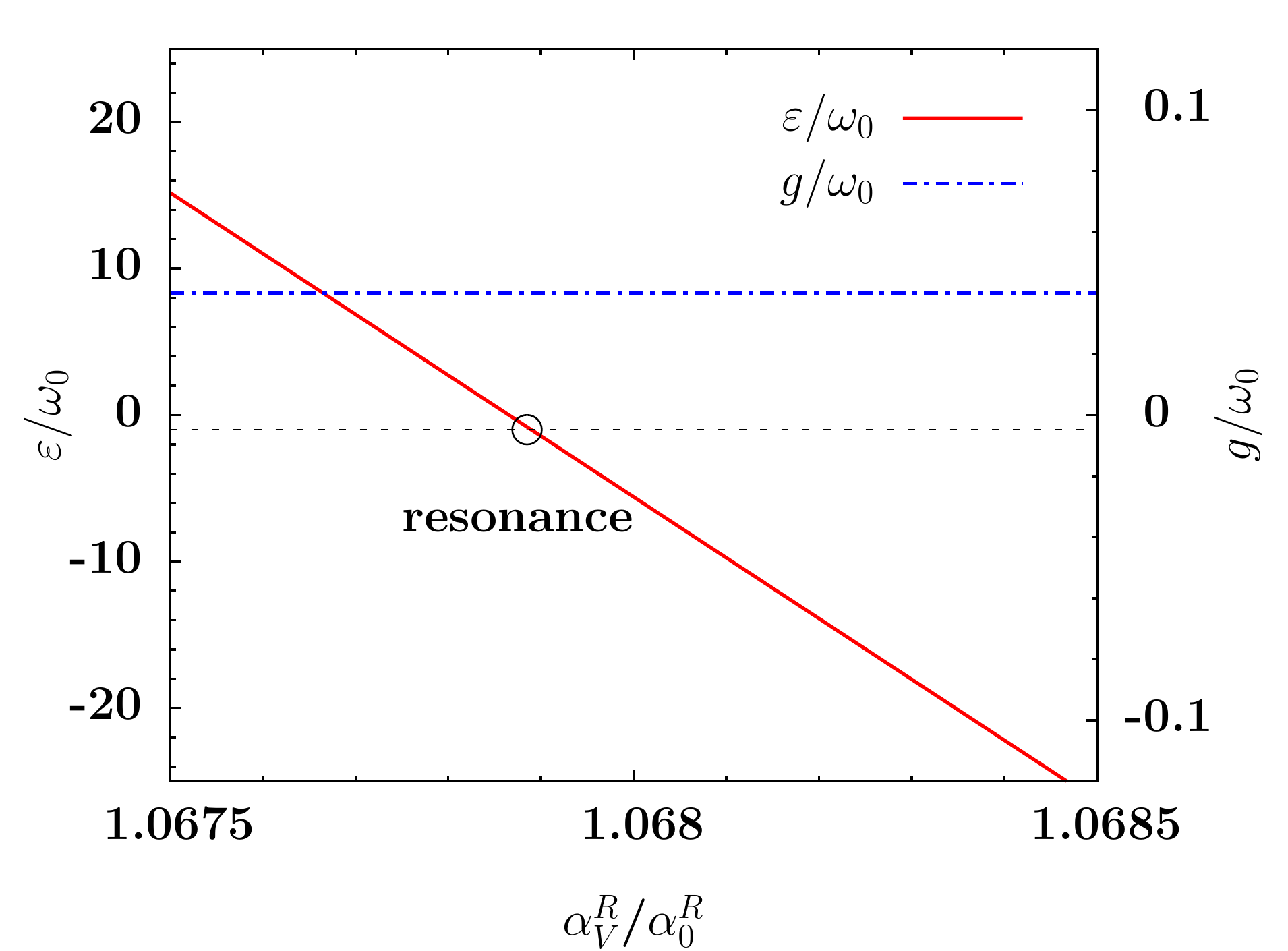}}
  \caption{(Color online) Qubit splitting $\varepsilon$ and the qubit-phonon
    coupling $g$ (the scale is on the right-hand side of the frame) versus
    $\alpha^{\rm R}_V$, the Rashba SOC strength in the topological ($T$) domains modulated by the
    gate voltage $V$.}
  \label{fig3}
\end{figure}

We now look into the qubit-phonon coupling and the resonance condition. We assume that the
nanomechanical oscillator FM2 has a mass $M=10^{-15}$~Kg and an oscillation frequency $\omega_0=5$~MHz. With
these parameters, the zero-point motion of the oscillator is calculated to be
${\tilde x_0}=0.1$~pm. We consider the symmetric case with
$l_{12}^0=l_{34}^0=150$~nm, and hence we have $\delta=0$ and
$g=0.2$~MHz in Eq.~(\ref{he}). The longitudinal length $l_{23}$ of the FM2 gate is chosen as
400~nm, such that the Rabi resonance condition
$\varepsilon \approx -\omega_0$ 
can be easily satisfied, e.g., by further subtly adjusting the gate voltage $V$
which controls the local Rashba SOC strength $\alpha^{\rm R}_V$. In Fig.~\ref{fig3}, we present the variation of $\varepsilon$ as
well as $g$ versus $\alpha^{\rm R}_V$. It is shown that when slightly adjusting $V$, and hence $\alpha^{\rm
  R}_0$, the resonance point $\varepsilon \approx -\omega_0$ can be reached while the
qubit-phonon coupling $g$ remains almost invariant. This qubit-phonon coupling is
relatively strong, in view of the long lifetime of the Majorana qubit and the high quality factor
of the nanomechanical oscillator. In principle, the qubit-phonon coupling can be
stronger when the domain length $l_{12}$ (as well as $l_{34}$) becomes smaller (refer to
Fig.~\ref{fig2}). However, if the two edge modes $\gamma_1$ and $\gamma_2$
(as well as $\gamma_3$ and $\gamma_4$) are too close and hence their
hybridization becomes quite strong, the model Hamiltonian (\ref{hm}) describing four distinguishable MBS
might fail.

\section{Qubit-phonon dynamics}
\label{dynamics}

Here we study the dynamics of the qubit-phonon hybrid system. To achieve
this, we make use of the Python-based Qutip software package\cite{johansson1760,johansson1234} to solve the Lindblad master equation,
\begin{align}\nonumber
  \dot\rho(t)=&-\frac{i}{\hbar}[H_{\rm eff},\rho(t)]+\frac{1}{2}\sum_{k}\Big\{[L_k,\rho(t)L_k^\dagger]\\&+[L_k\rho(t),L_k^\dagger]\Big\}.\label{master}
\end{align}
In this equation, $\rho$ is the density matrix of the qubit-phonon system, and $L_k$ are the Lindblad operators accounting for the dissipation of the hybrid
system due to its coupling to the environment. The relaxation of the Majorana
qubit is taken into account by $L_1=\sqrt{1/T_1}\sigma_-$, while the dissipation
of the nanomechanical resonator is included by $L_2=\sqrt{(\bar{n}+1)\omega_0/Q}a$
and $L_3=\sqrt{\bar{n}\omega_0/Q}a^\dagger$. Here ${\bar
  n}=[\exp(\hbar\omega_0/k_B{\tilde T})-1]^{-1}$ is the thermal phonon number in
equilibrium with the environmental temperature ${\tilde T}$, $Q$ is
the quality factor of the nanomechanical oscillator, and $T_1$ is the usual
relaxation time of the qubit. By solving the master equation, one can obtain the time evolution of the qubit and phonon occupations.

In our model, the temperature ${\tilde T}$ is set as 10~mK and hence the thermal phonon
number ${\bar n}$ is as large as 258. Therefore, an additional cooling
of the oscillator\cite{connell697,teufel359,chan89} is required, e.g., as also applied in
a proposed nanomechanical resonator--nitrogen-vacancy center hybrid system.\cite{arxiv1503.02437}
We assume that after side-band cooling\cite{connell697,teufel359,chan89} the phonons thermally occupy the lowest several quantum states
with a small phonon number, e.g., $n=0.3$. The initial state of the Majorana qubit is set as
$|\uparrow\rangle$, implying that a single electron is splitted into the $\gamma_1$ and
$\gamma_4$ Majorana fermions. Experimentally, this initial state might be realized when only
the FM1 and FM3 gates are in proximity to the nanowire before inserting the
middle FM2 gate. The relaxation time of the Majorana qubit depends on the
concrete set-up and environment. Following Refs.~\onlinecite{rainis174533} and
\onlinecite{schmidt085414}, we typically set $T_1$ around 100~$\mu$s.

\begin{figure}[t]
  {\includegraphics[width=9cm]{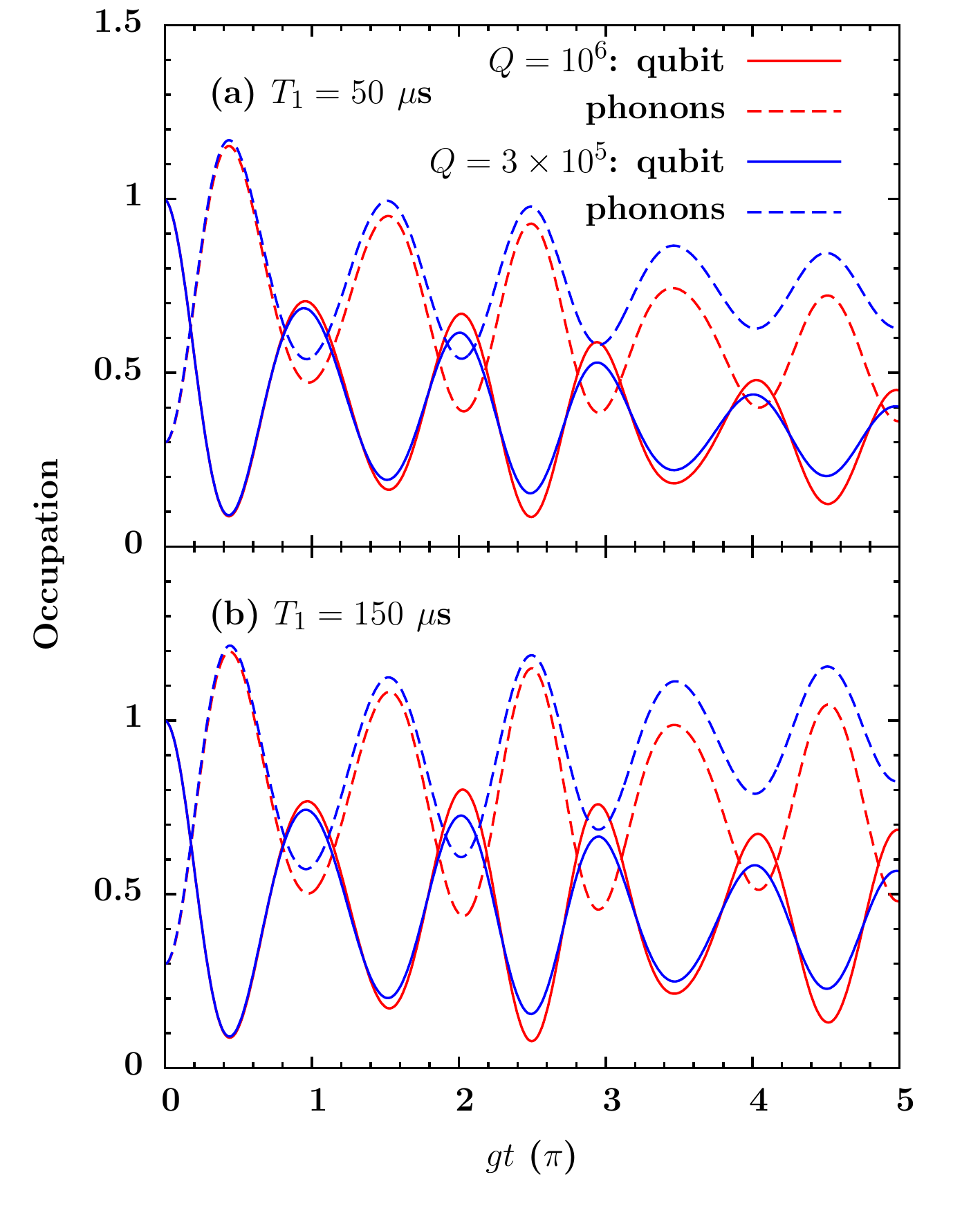}}
  \caption{(Color online) Time evolution of the occupations of the Majorana
    qubit (solid curves) and phonons (dashed curves) which are in Rabi
    resonance. The qubit relaxation time is set as 50~$\mu$s in (a) and
    150~$\mu$s in (b). The calculations for both (a) and (b) are performed with
    two different resonator quality factors: $Q=10^6$ and $Q=3\times 10^5$.}
  \label{fig4}
\end{figure}

In Fig.~\ref{fig4}, we plot the time  evolution of the occupations of the qubit
and phonons respectively, with different values of $T_1$ and $Q$. As indicated
by the figure, quantum information can be effectively transferred back and forth between the Majorana qubit and the
nanomechanical resonator.  During this process, the single electron in the
nanowire alternatively occupies (back and forth) the pair of MBS: $\gamma_1$ and $\gamma_4$, or $\gamma_2$
and $\gamma_3$. Inversely, this quantum information transfer
can also modulate the motion of the oscillator, e.g., the oscillation amplitude. In fact, as the nanomechanical resonator
is near its quantum ground state, the oscillation amplitude $\langle x_0^2\rangle$, which might be
observable, is almost linearly related to the phonon number. This is because $\langle
x_0^2\rangle\propto\langle(a^\dagger +a)^2\rangle\approx 2\langle a^\dagger
a\rangle +1=2n+1$. Therefore, the dashed curves in Fig.~4, representing the
time evolution of the phonon number, also supply information on the change of the
oscillation amplitude of the resonator due to its coupling with the qubit. This
phenomenon signifies the presence of a Majorana qubit. Certainly,
for better performance of this hybrid system (e.g., with a higher fidelity), a
higher quality factor of the resonator and a longer relaxation time of the qubit
are preferred.

\section{discussion}

Here we briefly compare our model to the one proposed by Kovalev {\sl
    et al.},\cite{kovaklev106402} where a vibrating cantilever is utilized to rotate a Majorana
  qubit. The effective Hamiltonian in their model [Eq.~(7) in Ref.~\onlinecite{kovaklev106402}] is in fact
equivalent to the one in our manuscript [Eq.~(5)]. This is understandable as both are in the framework of the spin-boson
model. Note that for both cases there exists a static off-diagonal term [for our case, that is the $\delta$ term in Eq.~(5)] coupling
the two levels of the qubit in the Hamiltonian. To neglect this term, in order to simplify the
theoretical analysis, some conditions have to be
satisfied. Specifically, in Ref.~5, a certain equilibrium angle ($\theta_0$
there) of the vibrating cantilever has to be established. In our opinion, exactly solving this angle and then
adjusting the experimental setup correspondingly\cite{kovaklev106402} are challenging. However, in order to neglect the constant off-diagonal
term in our case, the experimental setup must be mirror-symmetric about the
middle point of the FM2 gate, i.e., $l_{12}^0=l_{34}^0$. Therefore, we think
that our model is more easily accessible by experiments and hence more advantageous.

\section{conclusions}

In conclusion, we have proposed a hybrid system composed of a Majorana qubit and a mechanical resonator,
implemented by a semiconductor nanowire in proximity to an s-wave
superconductor. In this proposal, three ferromagnetic gates are placed on top of and along the
nanowire; the two outer gates are static and the inner one is free to oscillate harmonically as a mechanical resonator. These
ferromagnetic gates induce a local Zeeman splitting and
give rise to four Majorana bound states, constituting a Majorana qubit in the nanowire. The dynamical hybridization of
the Majorana bound states, arising from the motion of the oscillating gate,
results in a coherent coupling between the Majorana qubit and the mechanical
resonator.

This hybrid system can be adjusted to be in resonance,  e.g., with the assistance of a gate voltage on the
ferromagnetic gates, which controls the Rashba SOC locally in the nanowire. Our
study reveals that under resonance, a strong coupling between the qubit and the resonator can be
achieved.  Consequently, quantum information can be
effectively transferred from the Majorana qubit to the oscillator and then back
to the qubit. This quantum information transfer can manifest itself in
modulating the motion of the oscillator, which may conversely signify the
presence of the Majorana qubit.

\begin{acknowledgments}
The authors gratefully acknowledge X. Hu, G. Giavaras, L. Wang and Z. Li for valuable discussions and comments.
P.Z. acknowledges the support of a JSPS Foreign Postdoctoral Fellowship under Grant
No. P14330. F.N. is partially supported by the RIKEN iTHES Project, MURI Center
for Dynamic Magneto-Optics via the AFOSR award number FA9550-14-1-0040, the IMPACT program of JST, and a Grant-in-Aid for Scientific Research (A).
\end{acknowledgments}


\begin{thebibliography}{0}
\bibitem{note} Note that the pairing of MBS to form Dirac fermions
is arbitrary. A given pairing defines a basis and different bases can be
connected by a unitary transformation.\cite{nayak1083} Here we define Dirac
fermions according to the initial state where a single electron is splitted into
a pair of Majorana fermions $\gamma_1$ and $\gamma_4$ (refer to Sec.~\ref{dynamics}).
\end{thebibliography}
\end{document}